\documentclass[sigconf,authordraft]{acmart}
\renewcommand\footnotetextcopyrightpermission[1]{} % removes footnote with conference information in first column

\AtBeginDocument{%
  \providecommand\BibTeX{{%
    \normalfont B\kern-0.5em{\scshape i\kern-0.25em b}\kern-0.8em\TeX}}}

\setcopyright{acmcopyright}
\copyrightyear{2018}
\acmYear{2018}
\acmDOI{10.1145/1122445.1122456}

\begin{document}

\title{Incorporating Total Variation Regularization in the design of an intelligent Query by Humming system}

\author{Shivangi Ranjan}
\email{shivangiranjan14@gmail.com}
\affiliation{%
  \institution{Indian Institute of Technology Kanpur}
  \streetaddress{}
  \city{}
  \state{}
  \country{India}
  \postcode{208016}
}
\author{Vishal Srivastava}
\email{vishalsrivastava177@gmail.com}
\affiliation{%
  \institution{Indian Institute of Technology Kanpur}
  \streetaddress{}
  \city{}
  \state{}
  \country{India}
  \postcode{208016}
}

\renewcommand{\shortauthors}{}

\begin{abstract}
  A Query-By-Humming (QBH) system constitutes a particular case of music information retrieval where the input is a user-hummed melody and the output is the original song which contains that melody. A typical QBH system consists of melody extraction and candidate melody retrieval.

For melody extraction, accurate note transcription is the key enabling technology. However, current transcription methods are unable to definitively capture the melody and address inaccuracies in user-hummed queries. In this paper, we incorporate Total Variation Regularization (TVR) to denoise queries. This approach accounts for user error in humming without loss of meaningful data and reliably captures the underlying melody.

For candidate melody retrieval, we employ a deep learning approach to time series classification using a Fully Convolutional Neural Network. The trained network classifies the incoming query as belonging to one of the target songs.

For our experiments, we use Roger Jang’s MIR-QBSH dataset which is the standard MIREX dataset. We demonstrate that inclusion of TVR denoised queries in the training set enhances the overall accuracy of the system to 93\% which is higher than other state-of-the-art QBH systems.
\end{abstract}




%%  \includegraphics[width=\textwidth]{sampleteaser}
%%  \caption{Seattle Mariners at Spring Training, 2010.}
%%  \Description{Enjoying the baseball game from the third-base
%%  seats. Ichiro Suzuki preparing to bat.}

\maketitle
\pagestyle{plain} % removes running headers

\section{Introduction}
Recent times have witnessed the advent of network-based music fetching systems such as Spotify, iTunes etc. The emergence of such systems has led to a steep rise in creation of various audio databases. This calls for development of content-based retrieval methods that are well-adapted to specific characteristics of the audio data type [3]. Query-by-Humming is one such instance of a content-based retrieval system that can query an audio database with a user-hummed tune as input. A compelling music retrieval paradigm, QBH is positioned as the more natural alternative to standard text-based query systems in the music domain, since it does not require from the user any knowledge of the related metadata of the song  a priori.
Research in QBH is predominantly focused on addressing a) Melody extraction, and b) Candidate melody retrieval.

Any query or candidate melody can be analyzed to extract a sequence of notes or pitch values corresponding to each time instant. Melody contour is the overall shape of a song’s melody, when represented by a time series note sequence, also known as the pitch vector. Melody contour is one of the most important characteristics for identifying and cataloguing melodies. 
A typical QBH system bases its operation on effectively addressing inaccuracies found in user-hummed queries. Due to lack of user expertise and professional recording instruments, the user-hummed query has imperfections in addition to noise and distortion which is possible due to numerous factors, such as, stray mic clicks or improper recording, etc., leading to spurious peaks. Therefore, it becomes important to first smoothen the incoming query pitch vector, such that the system considers only the valid note transitions. It is to be noted that the same denoising process need not be done on melodies from the target database owing to the existence and accessibility of annotated representations (in the MIDI format) of the candidate melodies [16]. 
In this paper, we address the issue of smoothening the noisy queries with the help of Total Variation Regularization (TVR). Our approach hinges on the observation that denoising queries through Total Variation is more effective than basic approaches such as median filtering or linear smoothing because these methods do not preserve sharp edges. In contrast, TVR is notably effective at preserving edge information whilst smoothing out the noisy flat regions. Passing the input pitch vector through TV regularization eliminates the spurious peaks. In addition to TVR, the pitch vector is further processed through a slope filtering system which results in the final denoised, piece-wise constant pitch vector. This two-step process of melody extraction offers robustness against inaccuracies introduced in user-hummed queries.

Once the queries have been processed to remove the noise and become piecewise constant, they are included in the training set for our neural network architecture for candidate melody retrieval. The standard procedure of retrieving candidate melodies within a target database is through the combination of Locality Sensitive Hashing and DTW [9, 13] or through sequence alignment using edit-distance as a measure of similarity between two songs [11]. However, for the purposes of designing our QBH system with MIR-QBSH dataset, we have experimented with a deep learning approach. With the help of a Fully Convolutional Neural Network (FCNs), we perform classification of time series sequences in order to classify the incoming query as belonging to one of the target songs. We determine that with the inclusion of TVR denoised queries to train our neural network to perform time series classification, we are able to predict the target song with 93\% percent accuracy, which is higher than state-of-the-art systems.

\section{Related Work}
A seminal work in Query-By-Humming, Ghias et al. [6] produced a QBH system in 1995 by employing the approach of three methods for tracking pitch, namely, autocorrelation, maximum likelihood and cepstrum analysis, and found a modified version of autocorrelation to give the best performance. They then converted the pitch contour into a string sequence of Parsons Code composed of the alphabet {U, D, S}, for when the current note is higher than, equal to or lower than the previous note respectively. Furthermore, for candidate melody retrieval, they used a string alignment approach. Due to their dynamic-programming based pattern matching algorithm, the time taken to perform retrieval of songs was high.

In their QBH system, Ito et al. [7] aimed to address the issue of inaccurate F0 estimation resulting from automatic music transcription. The proposed system used multiple F0 candidates to absorb any possible error in estimating a single frequency, and the retrieval algorithm was a unique scoring function that took into account possible differences between a hummed query and the target song, namely, normalization of tempo and key matching.  It was observed from their experiments that considering multiple F0 candidates for note transcription is indeed leading to better overall performance of the retrieval system.

In order to address inaccuracies commonly found in user-hummed queries, Valero-Mas et al. [16] analyzed three different techniques of melody abstraction for encoding the estimated melodic pitch contours, namely, Symbolic Aggregate Approximation (SAX) which is an encoding for time series which provides both a dimensionality reduction as well as a lower bound in the distance calculations, PAA temporal segmentation with semitone quantization (PAA-ST) which is a modification of SAX replacing the statistical distribution approach to vertical quantization with a fixed grid of semitone quantization, and Pitch change segmentation with semitone quantization (PC-ST) which performs dynamic segmentation of melody contour whenever a pitch change occurs. The main takeaway is that although SAX is proven to have good performance for a range of time-series problems [18], it is not sufficient to design a QBH system, as demonstrated by the improvement in performance of the modified version PC-ST.

Ranjan et al. [11] proposed a QBH system that used the approach of Total Variation Denoising to extract melody from queries. This approach was founded on the fact that TVD effectively captures the underlying signal while performing noise reduction as it is designed to conserve the edge information of signals. For candidate melody retrieval, they followed a unique approach of incorporating semi-global sequence alignment techniques from the field of bioinformatics, which is a method to align a short sequence as a subsequence within a longer sequence. They evaluated their system on the MIR-QBSH dataset, and their system achieved an overall accuracy of ~71\%.

N. Mostafa et al. [9] proposed a combination of Convolutional Neural Network
(CNN) and Hidden Markov Model (HMM) based technique for transcription of notes. They used raw audio data as input to the CNN which learned the relevant feature vectors straight from raw audio data, bypassing the need for manual feature engineering. By employing Hidden Markov Models (HMM), they modeled the temporal aspect of note transcription. There were three states in the HMM, representing note transitions: the first state when the note is beginning, the middle steady note state and the last state when the note is decaying. Each of the HMM states had a posterior probability which was modeled by the CNN from the hummed raw audio sample. For candidate melody retrieval, they used Locality Sensitive Hashing (LSH) to return the closest approximate matches after which Dynamic Time Warping (DTW) and Earth Mover’s Distance (EMD) were used to compute the distance between each of those candidates. Overall, their QBH system was able to achieve an accuracy of ~92\%, which outperformed the existing approaches submitted to MIREX 2017. These experiments were performed on the MIR-QBSH dataset which is the standard MIREX dataset.

\section{METHODOLOGY}
\subsection{Dataset}
For training the our neural network to perform time series classification, we used two standard datasets: 
\begin{enumerate}
\item MIR-QBSH: This corpus [8] consists of user-hummed recordings of nursery rhymes collected over a period of 7 years from students in the “Audio Signal Processing and Recognition” course at the Computer Science Dept of Tsing Hua Univ., Taiwan. The corpus has been made available by Roger Jang and consists of two parts: 
\begin{itemize}
\item MIDI files: 48 ground truth monophonic MIDI files
\item WAVE files: 4431 wave file queries corresponding to the 48 MIDI songs. Each query is 8 seconds in length, and is recorded with a sampling rate of 8kHz.
\end{itemize}

\item MTG-QBH: This corpus [14] is composed of 118 sung melodies made by 17 different subjects, 9 female and 8 male, with a diverse range of musical experiences, from no experience to amateur musicians. The duration of the recordings varies, with the minimum length being 11 seconds to the maximum being 98 seconds, with 26.8 seconds being the average recording length.

\end{enumerate}
\begin{figure}[htbp]
  \centering
  \includegraphics[width=\linewidth]{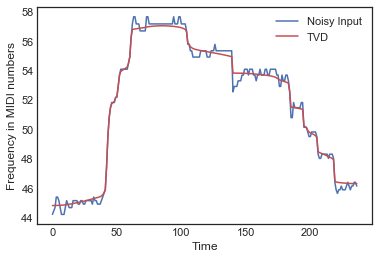}
  \caption{Illustration of noise removal by TVD while preserving edge information of the signal}
\end{figure}

\begin{figure}
  \centering
  \includegraphics[width=\linewidth]{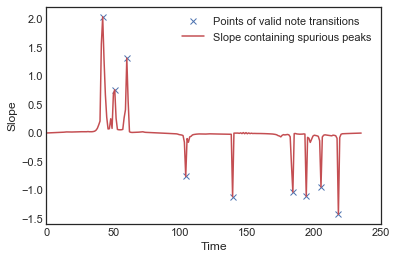}
  \caption{Illustration of slope filtering to identify points of sharp note transitions}
\end{figure}

\begin{figure}
  \centering
  \includegraphics[width=\linewidth]{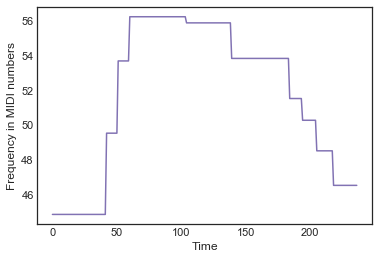}
  \caption{Reconstructed melody contour}
\end{figure}

\subsection{Melody extraction}
A robust process of melody extraction is an integral component
of any content retrieval system that targets music databases. As a first step, we extract the melody contour of each query using the implementation included in the Praat tool. Praat is a free software application designed for speech analysis in phonetics [20]. The transcribed note sequence extracted from Praat contains inaccuracies and melody distortion introduced during humming by the user. As a result, we need an efficient technique to remove background noise from the queries and identify the underlying melody. An approach for noise reduction used commonly in signal processing is Total Variation Regularization, proposed by Rudin et. al [12]. It is designed to preserve the edge information in a given noisy signal [15]. We use the approach as outlined in [11]. In contrast to the typical low-pass filter, TV regularization is described in terms of a problem of optimization, that is to say, it is equal to the L1 norm of a given signal’s gradient. This means that the foundation principle of TV Regularization is that an incoming signal that has an excessive and possibly spurious detail will display a large total variation, i.e., the absolute gradient of the signal will have a high integral value. In accordance with this principle, a reduction in the signal’s total variation will lead to the removal of undesired noise and spurious details whilst conserving necessary characteristics such as the edges within the signal. This approach of TV Regularization achieves better results than basic methods such as linear smoothing or median filtering since TV Regularization is notably effective at preserving edge information whilst denoising the noise in flat regions whereas these methods are not designed to preserve sharp edges, leading to loss of information about note transitions. Given an image i containing noise and other spurious details, we can employ the proposed noise reduction model to determine a corresponding image j over 2D space as the solution of a minimization problem [5, 11]:

$$\small{min \|u\|_{TV(\Omega)} + \frac{\lambda}{2} \int_{\Omega}^{} (f-u)^2 dx  , u \epsilon BV(\Omega)} $$

where, BV($\Omega$) = total variation over the domain, $\lambda$ = a penalty term.

The outcome of solving minimization of the aforementioned equation results in the desired denoised image. 

In the proposed QBH system, Total Variation Regularization is used as included in the versatile set of image processing routines of scikit-image [19], having extended it to 1-D time-series data, which is in this case, the hummed query. Hence, passing the input pitch vector through TV denoising eliminates the spurious peaks. 

However, the output is still not perfectly piece-wise constant as desired and thus, it is passed through an additional processing step. An additional slope filtering process is needed to reconstruct the input query into a piecewise constant signal. This filtering process identifies within the signal those instances of time where there is greatest probability of a note transition. By evaluating and examining the slope at each point of the smoothened pitch vector, the smaller peaks are removed and only the high magnitude slopes are retained. This is because only those slopes indicate a sharp note transition which have a high magnitude. In addition, if two peaks are close or overlapping, it means that two note transitions are occurring within a very short interval of time, which is not possible in a query hummed by a human. Hence, the local maxima is considered a valid transition. Finally, we are left with the indices of the slopes that correspond to valid note transitions. This new list of peaks is used to re-compute the final melody contour, which is piecewise constant, as desired. The steps followed to achieve melody extraction are shown in Figures 1, 2 and 3.

\begin{figure*}
 \centering
  \includegraphics[width=\textwidth]{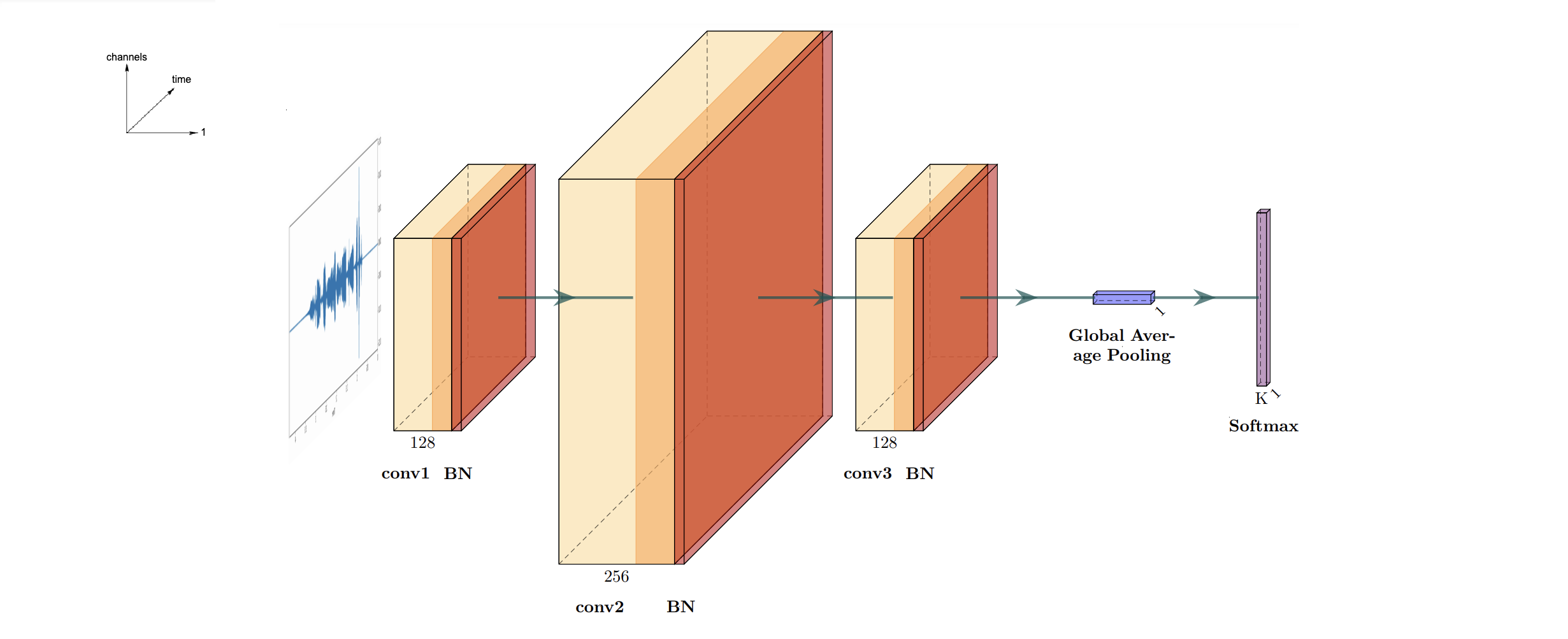}
  \caption{Fully Convolutional Neural Network architecture [17]}
  \Description{Illustration schematic showing the different layers used in an FCN}
\end{figure*}

\subsection{Candidate melody retrieval}
Once the queries have been appropriately processed, they are sent for candidate melody retrieval from the target database. Candidate melody retrieval is a key component of a QBH system. In this paper, we propose a FCNs-based candidate melody retrieval system. The basis of operation of our system is Time Series Classification (TSC) [1], which is a general area of Deep Learning that deals with how to assign labels to incoming time series which could be coming from one of numerous sources with the help of supervised learning. Design of a fast, accurate and scalable TSC system is essential and finds application across many subject-matter domains, such as retail advertising and cloud services [2]. A formal description of the problem statement in the context of QBH is as follows:

Given, ({\bfseries X}, {\bfseries y}) be the dataset, where {\bfseries X} represents the melody contour or time series representation of queries and {\bfseries y} is a discrete class variable, consisting of labels classifying each query into the correct target song. There are m observations of {\bfseries X} and n possible values of {\bfseries y}. The training dataset {\bfseries T} is made of k such training instances: ${\mathbf T} = \{ (X_1, y_1),..., (X_k, y_k) \}$. 

The problem of classifying the time series data comprises learning a classifier on {\bfseries T} to map from the input vector {\bfseries X} to a probability distribution over {\bfseries y}.

Fully Convolutional Neural Networks (FCNs) were first proposed in Wang et al. [17] for classifying time series data. FCNs do not include local pooling layers as a result of which the length of a time series remains unchanged across all convolutions. Another characteristic of FCNs is the use of Global Average Pooling (GAP) layer instead of a final FC layer used traditionally, which leads to a significant reduction in the number of parameters in the network. FCNs have been shown to be an effective approach to time series classification in domains such as document classification and speech recognition [4]. Our system employs FCNs to classify incoming queries as one of the target songs. The system architecture is shown in Figure 4. 

The proposed architecture consists of three convolutional blocks, followed by a GAP layer and a layer of traditional softmax classifier. Each convolution block performs three operations, that is, a convolution after which batch normalisation is done and the result is fed to ReLU activation function. The three convolution layers contain 128, 256 and 128 filters respectively and the lengths of the filters are 8, 5 and 3 respectively. To preserve the length of time series, each convolution has zero padding and a stride of one. The output of the final convolutional block goes into the GAP layer where it undergoes averaging over the entire time domain. Finally, the softmax classifier produces the desired output.

\section{EXPERIMENTS}
\subsection{Data preprocessing}
To train our FCN for candidate melody retrieval, we have primarily used the MIR-QBSH dataset. In addition to this dataset, we have also appended MTG-QBH dataset to the MIR-QBSH dataset to form our training set. This lends an increased number of data points as well as diversity to the training set, thereby leading to increased robustness to the training of our network. MIR-QBSH  dataset consists of 4431 queries and 48 target songs, while MTG-QBH dataset consists of 118 queries and 81 target songs. In the interest of increased accuracy of the trained network and uniformity within the database, each query is split into equal-length frames, with window size = 5 seconds and hop size = 2 seconds.

\subsection{Baseline System}
A fast and scalable baseline system has been designed to demonstrate the efficiency of the proposed retrieval method in juxtaposition. The architecture for the baseline is a simple Multi Layer Perceptron (MLP), composed of 5 layers with each layer fully connected to the previous layer [10]. The last layer of the network is a softmax classifier, fully connected to the output of its previous layer. The number of neurons in each hidden layer is 300, with activation function ReLU.

\subsection{Evaluation of the proposed QBH system}
Experiments were performed with two versions of the dataset mentioned in Section 3.1. First, the network was trained simply on the original noisy dataset, without the incorporation of TV Regularization. Later, the denoised version of queries was appended to our training set, which led to a drastic improvement in the retrieval accuracy, from 67\% to 93\%.

For training the FCN, initial learning rate was 0.005 which was left unchanged till 30 epochs, after which the learning rate was reduced by 0.7 at each epoch until cross-validation accuracy hit a plateau. 25\% of the dataset was used as a validation set for the network. We demonstrate that with the inclusion of denoised queries through TV Regularization in our dataset, for any given query, the proposed system of FCN returns the correct target song with 93\% accuracy, which is higher than the state-of-the-art system proposed by Mostafa et al. [9] as well as all other systems submitted to MIREX. Our system has the same accuracy as the powerful TYCX4. Although, it is to be noted that TYCX4 is a frame-based QBH system and is much more computationally expensive than our note-based system. The evaluation results are tabulated in Table 1. 

\begin{table}
  \caption{Evaluation of proposed QBH system with other state-of-the-art systems}
  \label{tab:freq}
  \begin{tabular}{ccl}
    \toprule
    Algorithm&Accuracy\\
    \midrule
    With TVR+ FCN (Proposed) & 0.93\\
    Without TVR+ FCN & 0.67\\
    With TVR+ MLP (Baseline) & 0.78\\
    Mostafa et al. & 0.92\\
    BS1 & 0.86\\
    WHLX1 & 0.47\\
    TYCX4 & 0.93\\
    ZH1 & 0.89\\
  \bottomrule
\end{tabular}
\end{table}

\section{DISCUSSION ANd future directions}
It is seen that a combined approach of denoising the input signal using TV Regularization and using FCNs for Time Series Classification on the resultant queries results in the highest accuracy of 93\%. Therefore, we conclude that our proposed system outperforms other state-of-the-art systems. This approach becomes especially powerful when the space of thousands of possible input queries maps onto a very small set of target songs. In these cases, employing FCNs to classify time series data proves to be an efficient solution to rapidly increase the scalability of the system, while also providing enhanced accuracy. Despite the enhanced performance, the proposed system has room for improvement. Although note-based QBH systems are more efficient than frame-based QBH systems in that they involve relatively uncomplicated feature engineering and are faster while frame-based systems tend to be more computationally expensive, yet, a frame-by-frame analysis of the input query and the target song yields a much richer source of information and consequently, improves the retrieval accuracy. The accuracy can be further enhanced by considering additional features such as velocity with which each note is hummed, and other modifications in feature engineering.

Despite numerous scholars working in the field of Query-by-Humming, this field is still under active development. There are various approaches yet to tested out in both components, namely, melody extraction and candidate melody retrieval.
\begin{enumerate}
    \item An important step would be the design of a system that performs automatic music transcription from raw audio data with multiple candidates of fundamental frequency. This approach offers a two-fold advantage of absorbing any errors in identifying a single frequency by giving multiple candidates, as well as aid in the design of an end-to-end QBH system.
    \item Crucial insight is to be gathered and accuracy to be further improved by visualising songs in the frequency domain. For instance, chroma features are widely known to have superior performance in audio cover song identification tasks. They are, however, yet to be explored in the context of QBH systems.
    \item Spectral analysis of songs will not only provide information that would otherwise be lost in the time series representation, but also allow developers to leverage powerful pattern recognition methods arising from research in Computer Vision and be able to apply them in the music domain. Such an interdisciplinary approach holds promise of bringing the two fields closer.
\end{enumerate}

\section{Conclusion}
In this work, we have proposed an efficient and intelligent Query-by-Humming system that incorporates Total Variation Regularization in its design, resulting in enhanced accuracy and performance of candidate melody retrieval. For the latter, our system utilizes a Fully Convolutional Neural Network which has proved to be effective in performing Time Series Classification. Consequently, our system achieved an enhanced accuracy of 93\% which is better than other state-of-the-art systems. It bears mentioning that a salient feature of FCNs is the invariance in the number of parameters used in each layer except the last for time series of variable lengths. Therefore, it enables the possible use of transfer learning since one can train the model with a certain dataset after which it can be further honed on a different target dataset. This allows us to circumvent the necessity for a large amount of new data and results in improved performance of the overall neural network.

\end{document}